\documentclass[structabstract]{aa}  
\usepackage{graphicx}
\usepackage{txfonts}
\usepackage[]{natbib}  
\usepackage[toc,page]{appendix}

\usepackage{epstopdf}

\newcommand\kms{\ensuremath{\mbox{km}\,\mbox{s}^{-1}} }
\newcommand\Teff{\ensuremath{T_\mathrm{eff}}}
\newcommand\logg{\ensuremath{\log g} }
\newcommand\FeH{\ensuremath{\mathrm{[Fe/H]}} }
\newcommand\TiFe{\ensuremath{\mathrm{[Ti/Fe]}} }
\newcommand\NiFe{\ensuremath{\mathrm{[Ni/Fe]}} }
\newcommand\XFe{\ensuremath{\mathrm{[X/Fe]}} }

\begin{document}
\title{SPADES: a stellar parameters determination software}

\author{H.~Posbic\inst{\ref{inst1}}\and  D.~Katz\inst{\ref{inst1}}\and
E.~Caffau\inst{\ref{inst2},}\inst{\ref{inst1}}\and P.~Bonifacio\inst{\ref{inst1}}
\and A.~G\'omez\inst{\ref{inst1}} \and L.~Sbordone\inst{\ref{inst2},}\inst{\ref{inst1}}
\and F.~Arenou\inst{\ref{inst1}}}

\institute{GEPI, Observatoire de Paris, CNRS, Universit\'{e} Paris Diderot, Place Jules Janssen, 92190, Meudon, France\label{inst1}
\and
Zentrum f$\ddot u$r Astronomie der Universit$\ddot a$t Heidelberg, Landessternwarte, K$\ddot o$nigstuhl 12, 69117, Heidelberg, Germany\label{inst2}
}

   \date{Received 14 April 2012 / Accepted 16 July 2012 }
  
 
  \abstract{
As increasingly more spectroscopic data are being delivered by medium- and high-resolving power multi-object spectrographs, 
more automatic stellar parameter determination softwares are being developed.
The quality of the spectra collected also allows the determination of elemental abundances.}
{SPADES is an automated software for determining: the radial velocity (Vr), the effective temperature (\Teff), the surface gravity (\logg),
the metallicity (\FeH), and most importantly, the individual abundances. In this first version it is targeted on the analysis of mid-F-G dwarfs, but is
meant to evolve to analyze any type of single stars.}
{SPADES relies on a line-by-line modeling to determine the stellar parameters.}
{The internal systematic and random errors of SPADES were assessed by Monte Carlo method simulations
 with synthetic spectra and the external systematic errors by analysing real ground-based observed spectra.
For example, by simulating the Giraffe setups HR13 and HR14B with synthetic spectra for a dwarf
with $\Teff=5800$~K, $\logg=4.5$, $\FeH=0.0$~dex and with a signal-to-noise ratio (snr) of 100, the
stellar parameters are recovered with no significant bias and with 1-$\sigma$ precisions of 8 K for \Teff,
0.05 for \logg, 0.009 for \FeH, 0.003 for \TiFe and 0.01 for \NiFe.}
{}

\authorrunning{H. Posbic et al.}
\titlerunning{SPADES}
\maketitle

%

\section{Introduction}
The present and future large spectroscopic surveys are going to significantly increase the number of spectroscopic
data to be analyzed. A few examples are RAVE with some 250~000 stars observed so far \citep{RAVE11,Siebert11},
the Gaia-ESO Survey with about $10^5$ stars to be observed, Gaia with
about 200 million stars \citep{Katz04,Katz09}. To analyze these quantities of data, automatic
spectra analysis softwares are needed.\\

The development of automated spectra analysis software can be traced back to the early 1990s.
Since then, many programs were developed based on a wide variety of methods: e.g.
orthogonal vector projection \citep{Cayrel91, Perrin95}, minimum distance method,
e.g. TGMET \citep{Katz98, Soubiran03}, and ETOILE \citep{Katz01}, 
artificial neural networks \citep{Bailer00, AllendePrietoEtAl2000}, feature fitting softwares such as Abbo
\citep{Bonifacio03}, projection vector softwares such as MATISSE
\citep{A.Recio-Blanco06, BijaouiEtAl2008},
principal component analysis softwares such as $MA\chi$ \citep{Jofre10},
and forward modeling algorithm ILIUM \citep{Bailer10}. One can also cite DAOSPEC
\citep{StetsonPancino2008, StetsonPancino2010} and FITLINE \citep{FrancoisEtAl2003},
which automatically measure equivalent widths. DAOSPEC and FITLINE are not automated per-say
parameter determination softwares, but coupled with programs that derive atmospheric parameters
from equivalent widths, they are used by many astronomers to automatically parameterize stars.\\

The work presented here describes the development of a new automatic stellar spectra analysis software.
This is based on line-by-line analysis. The reference lines are modeled.
One of the motivations to develop SPADES (Stellar PArameters DEtermination Software) is the analysis of a set of 200 Giraffe HR13 and HR14
spectra, mainly F-G dwarfs, that were collected in the context of a study of the structure of the Galactic disk.
Therefore, the first version of SPADES is illustrated and tested with simulated
Giraffe-like spectra of mid-F-G dwarfs. SPADES is coded with Java.\\

SPADES is described in Sect.~\ref{author:sect1}. It relies on auxiliary data: a line list and a grid of
reference spectra. They are presented in Sect.~\ref{author1:sect2}. The performances of SPADES were
evaluated by Monte Carlo method simulations against synthetic spectra and by comparison to observed
spectra of the Sun, $\nu$~And, $\beta$~Vir, $\mu$~Her, and $\sigma$~Dra
(see Sect.~\ref{author1:sect3}).


\section{SPADES}
\label{author:sect1}
	\subsection{General concept}
        SPADES automatically determine: the radial velocity (Vr), the effective temperature (\Teff),
	the surface gravity (\logg), the metallicity (\FeH), and individual abundances (\XFe for the element X)
	for single stars. This first version of SPADES is focused on mid-F-G dwarfs and the micro-turbulence
        is set to 1~\kms.\\

	Many of existing automatic spectra analysis softwares uses global methods, meaning they process all pixels of the 
        studied spectrum \citep{Katz98, Katz01, Soubiran03, A.Recio-Blanco06}.
	With SPADES we aimed to explore another possibility, which is the line-by-line analysis. 
	Each parameter is determined using a pre-defined set of lines. 
	The merits of this technique are, first, the possibility to select lines that are particularly sensitive to the stellar parameter to be determined,
	and second, this allows one to exclude lines that are not correctly modeled.
	A drawback is that only a fraction of the total available information is used.

	Often, whether automated or manual, methods using a line-by-line analysis 
	rely on equivalent width measurement and use a curve-of-growth analysis. 
	We chose to rely on synthetic spectra modeling and profile fitting. 
	The motivation is twofold. First, this takes care of small blend problems. Assuming
        that the blend is correctly modeled, its contribution to the observation
        and to the modeled profile will (partly) cancel out. Second, some parameters
        can be derived by line profile fitting, e.g. \Teff from the wings of the Balmer
        lines. A drawback of this method is the need for a very large library of synthetic spectra.
	Reading and processing many synthetic spectra is computationally demanding and
        therefore the analysis is slower.

	In the line-by-line analysis the accuracy and precision of the results 
	is very sensitive to the determination of the continuum or pseudo-continuum. 
	Yet, this is often difficult particularly in cool and/or super metal-rich stars. 
	SPADES determines ``local'' continuum from the comparison of the reference
        (synthetic) spectra and the studied spectrum (see sect.~\ref{author1:subsect1}).

	As opposed to mean metallicities, SPADES determines iron and individual elements
        abundances. To derive of the individual abundances SPADES calculates
        on-the-fly atmospheric models and synthetic spectra for the values of \Teff,
        \logg, \FeH it had determined (see sect.~\ref{IndividualAbundances}).

	SPADES can simultaneously analyze various spectral domains (possibly with
        different resolving powers) of the same target, e.g. different Giraffe setups,
        or different orders of an echelle spectrograph, or even spectral domains collected
        with different spectrographs. 

	\subsection{Processing overview}

	A SPADES analysis proceeds in three steps: it first determines the radial velocity, 
	then the atmospheric parameters, and finally it determines the individual abundances.\\

	The radial velocity is determined by cross-correlation with a synthetic template (see sect.~\ref{RadialVelocity}).\\

	The three atmospheric parameters (effective temperature, surface gravity, iron over hydrogen ratio) are determined independently 
	(see sect.~\ref{Parameters}). 
	Currently, each parameter is associated to one diagnostics. 
	SPADES was conceived to aggregate more diagnostics along its life span and more will be added in the future
	Current diagnostics are
        \begin{itemize}
         \item effective temperature: Balmer line wings fitting (suited for \Teff derivation in F-G dwarfs),
	 \item surface gravity: ionization equilibrium,
	 \item iron abundance: neutral iron line profile fitting.
        \end{itemize}

        The individual abundances are determined element per element by line profile fitting.\\

	The reader will notice that some of these diagnostics are similar to those used in the so-called ``detailed analysis''. 
	Two differences should be emphasized: (i) SPADES relies on profile fitting and modeling rather
        than equivalent width measurement, and (ii) the SPADES precision is not limited by the step in the grid of
        atmospheric models and spectra (see~Sect.~\ref{Overview}).

	  \subsection{Pre-processing}
          \label{Preprocessing}
	  Each of the three processing steps includes pre-processing tasks.
	  Pre-processing a spectrum includes shifting to rest frame, excluding cosmics, excluding telluric lines, convolving the synthetic spectrum
          to the resolving power of the ground-based observed spectrum, normalizing to continuum and excluding possible unexposed edges of the spectrum. 
	  Every functionality remains an option: it can be switched on and off in the configuration file by the user.\\

	  Shifting to rest frame is performed by using either a radial velocity provided by the user or that 	
	  determined by SPADES (see sect.~\ref{RadialVelocity}).

	  The exclusion of cosmics is performed by detecting the fluxes that lie significantly above the continuum and
          replacing their value by interpolation with the neighboring pixels.

	  The wavelengths of the telluric lines were pre-tabulated. These lines are not cut. 
	  Their pixels are simply flagged as not valid and are excluded from all subsequent analysis. 

          The spectral resolution of the synthetic spectra should be matched to the resolution of the ground-based observed spectrum. 
	  To do so, the reference spectra are convolved using a Gaussian profile (valid for slow rotators). 

          \begin{figure}[ht!]
	  \includegraphics[width = 270 pt]{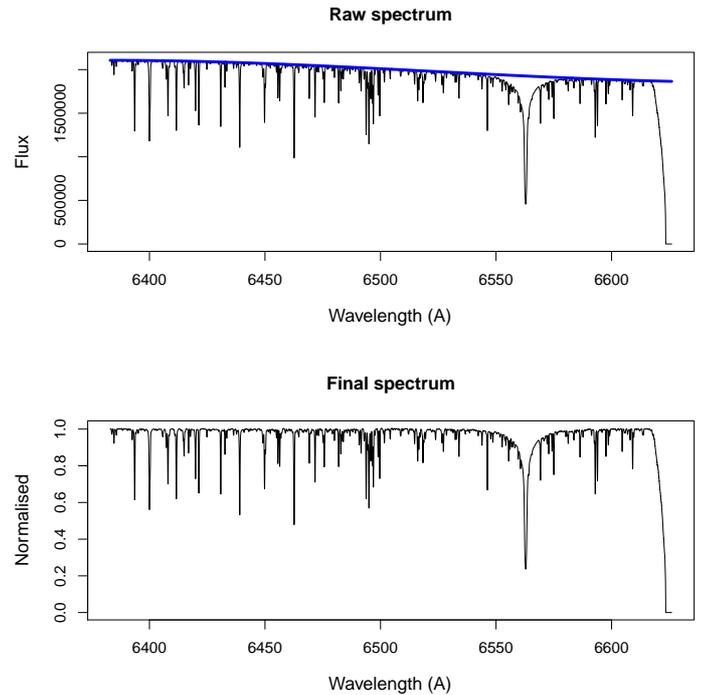}
	  \caption{Observed Giraffe/VLT (HR14B setup) Sun spectrum before normalization, with the polynomial fit shown as a
          thick blue line (top) and after normalization (bottom)} 
	  \label{author1:fig1}
	  \end{figure}

	  The synthetic spectra are already normalized to continuum. 
	  The studied spectrum is iteratively fitted with a polynomial function, using a sigma-clipping technique
          to exclude the lines and cosmic rays from the fit. The degree of the polynomial is defined by the user,
          depending on the large-scale shape of the spectrum. The studied spectrum is normalized to 1 at the level of the continuum
          by division by the polynomial. Figure \ref{author1:fig1} presents an observed Giraffe/VLT (HR14 setup) Sun
          spectrum before normalization, with the polynomial shown as a thick blue line (top) and after normalization
          (bottom). 

	  Edge exclusion is necessary if the ground-based observed spectrum suffers from underexposed or unexposed pixels.
          SPADES can automatically detect and flag as not valid the underexposed/unexposed pixels.

\subsection{First step: Radial velocity}
\label{RadialVelocity}
The radial velocity determination is based on a cross-correlation of the studied
spectrum and a synthetic template spectrum.\\

The atmospheric parameters of the template
spectrum are chosen by the user.
 At this stage in the processing, SPADES has not yet provided estimates for the
atmospheric parameters of the studied source. If no external information is available
to guide the choice of the template, it is likely that there will be a mismatch between
the ground-based observed spectrum and the selected template. Monte Carlo method simulations were performed
to assess the impact of selecting a template 400~K and 1~000~K warmer than the studied
spectrum. The tests were conducted with 200 snr=50 synthetic studied spectra of solar parameters.
In both cases (400 and 1~000~K offsets), the mismatch leads to a bias of about 0.2~\kms
and no significant change in the precision of the estimated radial velocities (see
table~\ref{table:1}). A bias of 0.2~\kms represents about 1/60$^{th}$ of the full width at
half maximum (FWHM) and has a negligible impact on the following steps of the analysis.
If, a posteriori, the parameters estimated by SPADES and those of the selected template
are significantly different, the radial velocity can be re-estimated with a better matching
template.\\

The studied and the template spectrum are pre-processed. 
The reference spectrum is then shifted step by step in radial velocity over an interval
defined by the user. In the tests, we used an interval of -500 to 500 km/s. At each step,
the synthetic spectrum is re-sampled to the same sampling as the studied spectrum. At each
step, the correlation coefficient of the two spectra is calculated:
	
\begin{equation}
CC(\mathrm v) = \frac{1}{N}  \sum^{N}_{i = 0} \frac{(S(i) - <S>) \times (R(i) - <R>)}{stdv(S) \times stdv (R)},	 
\end{equation}

where $\mathrm v$ is the velocity shift applied to the template spectrum, $N$ is the number of valid pixels in the studied spectrum,
$S$ (resp. $R$) the studied (resp. reference) spectrum counts, $<S>$ (resp. $<R>$) the mean 
studied spectrum (resp. reference) counts, and $stdv$ the standard deviation: 

\begin{equation}
stdv(X) =  \big(\frac{1}{N}{\sum^{N}_{i = 0} (X(i)-<X>)^{2}}\big)^{1/2},
\end{equation}

\begin{figure}[ht!]
		  \includegraphics[width = 270 pt]{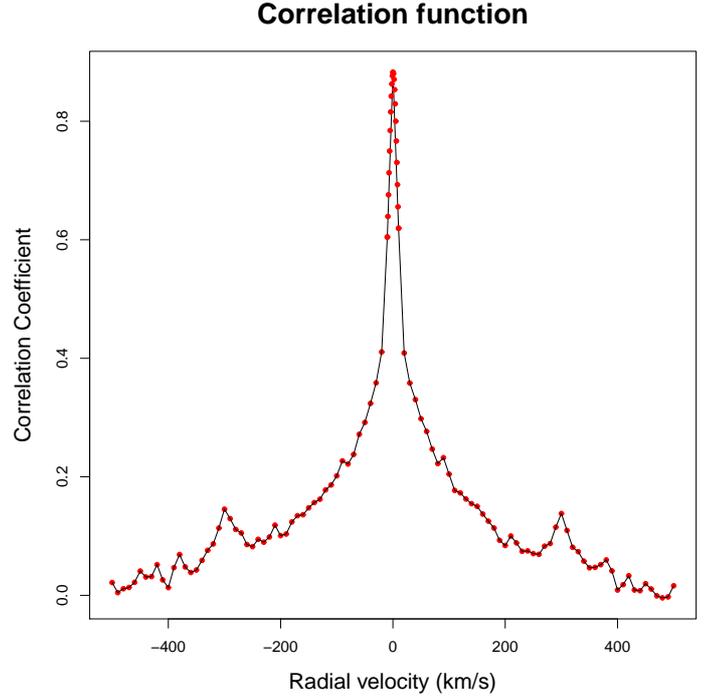}
		  \caption{Correlation function for a synthetic studied spectrum
        of solar parameters (\Teff~=~5777~K, \logg~=~4.44 and \FeH~=~0.0~dex) and S/N~=~100 (Poisson noise) and a noiseless
        synthetic template with the same atmospheric parameters} 
		  \label{author1:fig2}
		  \end{figure}

	The correlation coefficients as a function of the radial velocity shifts form the cross-correlation function. 
	Figure~\ref{author1:fig2} shows the cross-correlation function obtained for a synthetic studied spectrum
        of solar parameters (\Teff~=~5777~K, \logg~=~4.44 and \FeH~=~0.0~dex) and S/N~=~100 (Poisson noise) and a noiseless
        synthetic template with the same atmospheric parameters. 

	The radial velocity of the observed source corresponds to the maximum of the correlation function.
        The maximum is derived in two steps. First, the highest correlation coefficient is found and then
        a second-degree polynomial is adjusted to the correlation function in the vicinity of the highest
        coefficient. This allows one to estimate the radial velocity with a precision better than the step used
        to shift the template.\\ 
	
	The error on the radial velocity determination, $\sigma_{Vr}$ is estimated with the \cite{Zucker03} formula:
	\begin{equation}
	 \sigma_{Vr}^2 = - \Bigg[ {N\times\frac{C''(m)}{C(m)}\frac{{C(m)}^2}{1-{C(m)}^2}} \Bigg] ^{-1},
	\end{equation}
	where $C(m)$ is the value of the correlation function at its maximum, $C''(s)$ the value of the second derivative of the correlation function at the maximum
        of the correlation function and $N$ the number of valid\footnote{Some non-valid pixels affected by, e.g. telluric lines, could be excluded from the derivation
        of the correlation function (see Sect.~\ref{Preprocessing}).} pixels in the ground-based observed spectrum.\\

        The performances of the radial velocity estimation of the radial velocity were assessed with a Monte Carlo method.
	For three S/N: 30, 50 and 100, Giraffe HR13 and HR14B spectra were simulated (200 per SNR) by injecting a synthetic spectrum
        with Poisson noise. The synthetic spectrum was a solar-like spectrum in the rest frame. The same spectrum, but noiseless, was used as template.
        Table~\ref{table:1} presents the median and dispersion of the distributions of radial velocities derived by SPADES for each SNR.
        Because the true radial velocity of the studied spectra is 0~\kms, the median of a distribution is an estimate of the systematic error
        and the dispersion an estimate of the random error.

\begin{table}[h!]
\caption{Median and dispersion (in \kms) of the radial velocity distribution derived by SPADES for 200 studied
spectra for SNR 30, 50 and 100.}
\label{table:1}
\centering       
\begin{tabular}{c | c c}
SNR & Vr median & $\sigma$ Vr \\
    &  (\kms)   & (\kms)      \\ \hline
100 & -0.005    &  0.03       \\
 50 & -0.006    &  0.06       \\
 30 &  0.002    &  0.11       \\
\end{tabular}	
\end{table}

        \subsection{Second step: Stellar parameter determination}
	\label{Parameters}
        \subsubsection{Overview}
	\label{Overview}
	SPADES determines the effective temperature (\Teff), the surface gravity (\logg), and the iron abundance (\FeH).
        In this first version targeted on mid-F-G dwarfs, we assumed a micro-turbulence of 1~\kms for all our stars,
        including the tests with the observed spectra.

	The determination of the atmospheric parameters is iterative. Initial values for the three parameters are provided
	either by the user or by SPADES (see Sect.~\ref{TGMET}). 
	Each parameter is determined independently of the two others, leaving these two at their initial values. 
        At the beginning of a new iteration, the values of the three parameters are updated, therefore providing new ``initial'' values.	
	This is repeated until the convergence condition is met: i.e. absolute differences between the parameters obtained at iteration
        $n$ and $n-1$ all simultaneously lower than 10~K for \Teff, 0.1 for \logg, and 0.025 for \FeH.
	The iterative approach was chosen over a direct approach (e.g. scanning the whole grid of 2000 reference spectra over all parameters at once),
        because it was computationally faster. Yet, the iterative approach could potentially be more complex in terms of convergence. The convergence
        performances of SPADES are assessed in Sect.~\ref{ConvergenceProperties}.\\
	
	At any iteration, the estimation of each parameter follows the same logic:\\ 
	SPADES first calculates a 1D grid of synthetic spectra, varying only along the parameter that is estimated. This 1D-subgrid
        is calculated from the large grid of synthetic spectra described in Sect.~\ref{referenceGrid}. The grid is linearly interpolated to the
        fixed values of the other parameters. As an example, at the start of iteration n, the starting values for the 
        atmospheric parameters derived from iteration n-1 are $\Teff = 5743 K$, $\logg = 4.42$, and $\FeH = -0.02$. The first 
        parameter to be determined in iteration n is \Teff. A 1D-subgrid is linearly interpolated for \Teff ranging from 5200 to
        6200 K by step of 200 K and $\logg = 4.42$ and $\FeH=-0.02$~dex. 
        
        Each synthetic spectrum is then compared (not globally, but the selected lines) to the ground-based observed spectrum
        according to the diagnostic appropriate for the parameters to be determined. As an example for \Teff, the 
        $H\alpha$ profile of the ground-based observed spectrum will be compared to that of each synthetic spectrum.

        The result of the comparison is a similarity indicator for each node in the 1D-subgrid.
        This series of similarity indicators as a function of the parameter to be determined form
        a function that is analyzed (minimized or nulled, depending on the function) for the best similarity
        value, leading to the best estimate of the parameter to be determined. Interpolation or polynomial
        fit allows one to determine the parameter with a precision that is not limited by the step
        of the 1D-subgrid.\\

 SPADES is iteratively searching for the best matching synthetic spectrum
in a 3D space (\Teff, \logg and \FeH). In this first version of SPADES,
the search is split in three 1D minimizations, rather than using a
multi-dimensional search method. The motivation is that this facilitates
visualizing and interpreting the residuals and similarity functions
(See e.g. Fig.~\ref{author1:fig8}) that depend on a single parameter.
In future versions, it would be interesting to experiment with multi-dimensional
minimization algorithms such as Gauss-Newton, gradient, or Levenberg-Marquardt methods.

\subsubsection{Starting parameters}
\label{TGMET}
 SPADES proceeds iteratively to derive the atmospheric parameters.
It therefore needs starting values for the effective temperature, surface
gravity, and iron over hydrogen ratio prior to the first iteration.
Those can be deduced from photometric or bibliographic information if
available.

If no external information is available, SPADES offers the
possibility to perform a first parameterization of the studied source,
using the TGMET/ETOILE method \citep{Katz98, Katz01, Soubiran03}.
The results of this first analysis are used as starting values for
the iterative parameterization with SPADES.

The TGMET/ETOILE method is direct (not iterative). It compares a studied
spectrum to a library of reference spectra, looking for those most similar
in the least-squares-sense. The parameters deduced for the studied spectrum
could either be the parameters of the most similar spectrum or a combination
of the parameters of the most similar spectra. Because SPADES is only looking
for starting values, it simply adopts the parameters of the most similar
spectrum.

The performances of the SPADES implementation of the TGMET/ETOILE method
(hereafter referred to as SPADES-TE)) were assessed with a Monte Carlo method method. Starting
from a synthetic spectrum with \Teff~=~5800~K, \logg~=~4.5, \FeH~=~$-$1~dex
(which is a node of the grid of synthetic spectra), three series of 200 studied
spectra each were simulated for S/N=100, 50 and 30, respectively, and the
resolving power and wavelength range of the Giraffe HR13 and HR14B setups.
At S/N=50 and 100, the correct node in the grid is recovered in 100\% of
the cases. At S/N=30, SPADES-TE returns most of the time a solution biased
by one grid step in \Teff (i.e. +200~K) and two grid steps in \FeH
(i.e. +0.4~dex).

SPADES-TE was also tested on a Sun spectrum observed with Giraffe HR13
and HR14B setups in twilight conditions. SPADES-TE found \Teff~=~5800~K,
\logg~=~3.5 and [Fe/H]~=~$-$0.2~dex, to be compared to the canonical
parameters for the Sun \Teff~=~5777~K, \logg~=~4.44, and \FeH~=~0~dex.\\ 

We recall that the version of SPADES presented here is
limited by the GRID of synthetic spectra and by the list of
reference lines to the mid-F-G dwarf stars. Future versions of the
program will rely on larger grid(s) and on several fine-tuned line lists,
which will extend the scope of SPADES.

\begin{figure}[ht!]
\includegraphics[width = 270 pt]{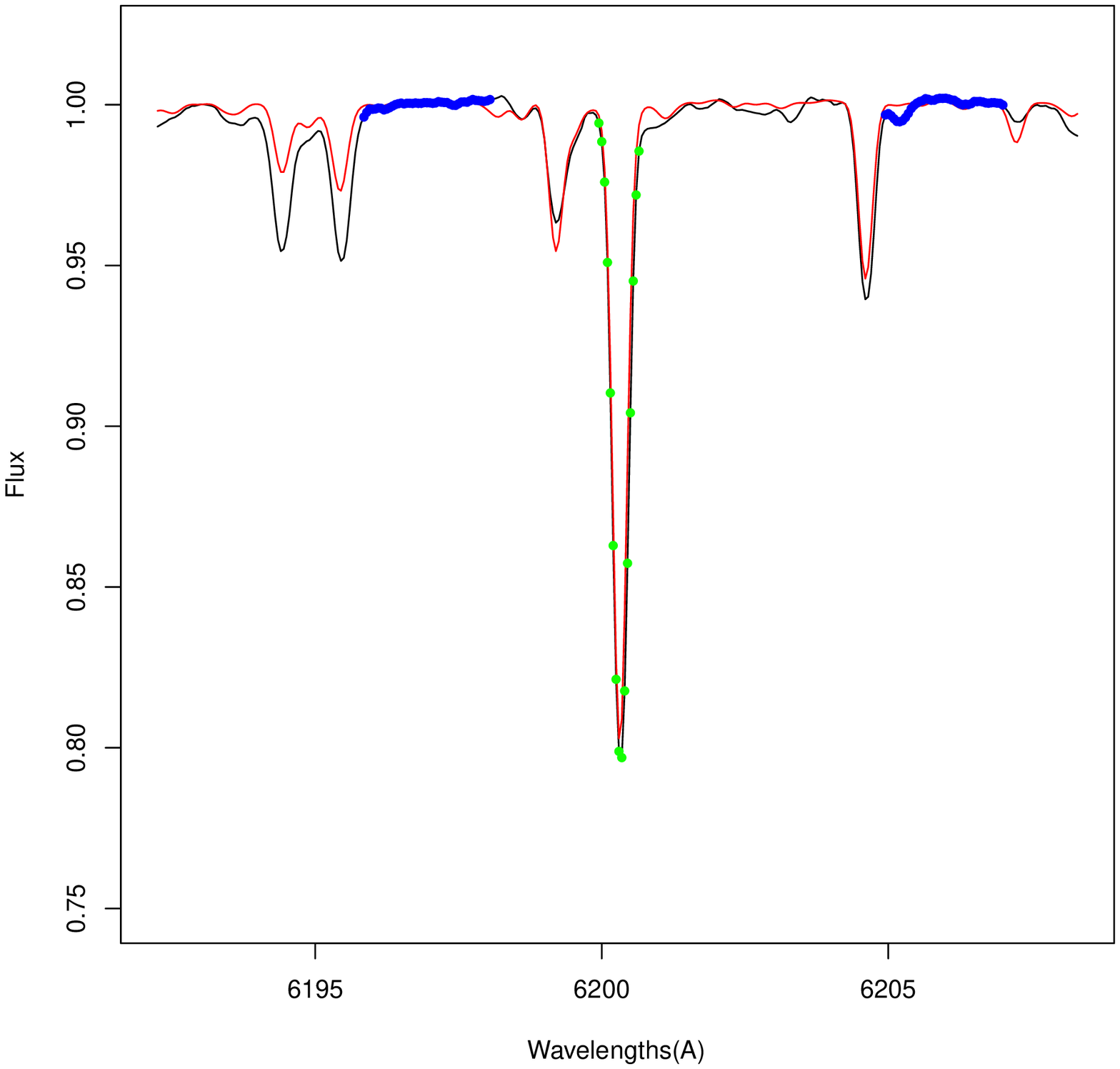}
\includegraphics[width = 270 pt]{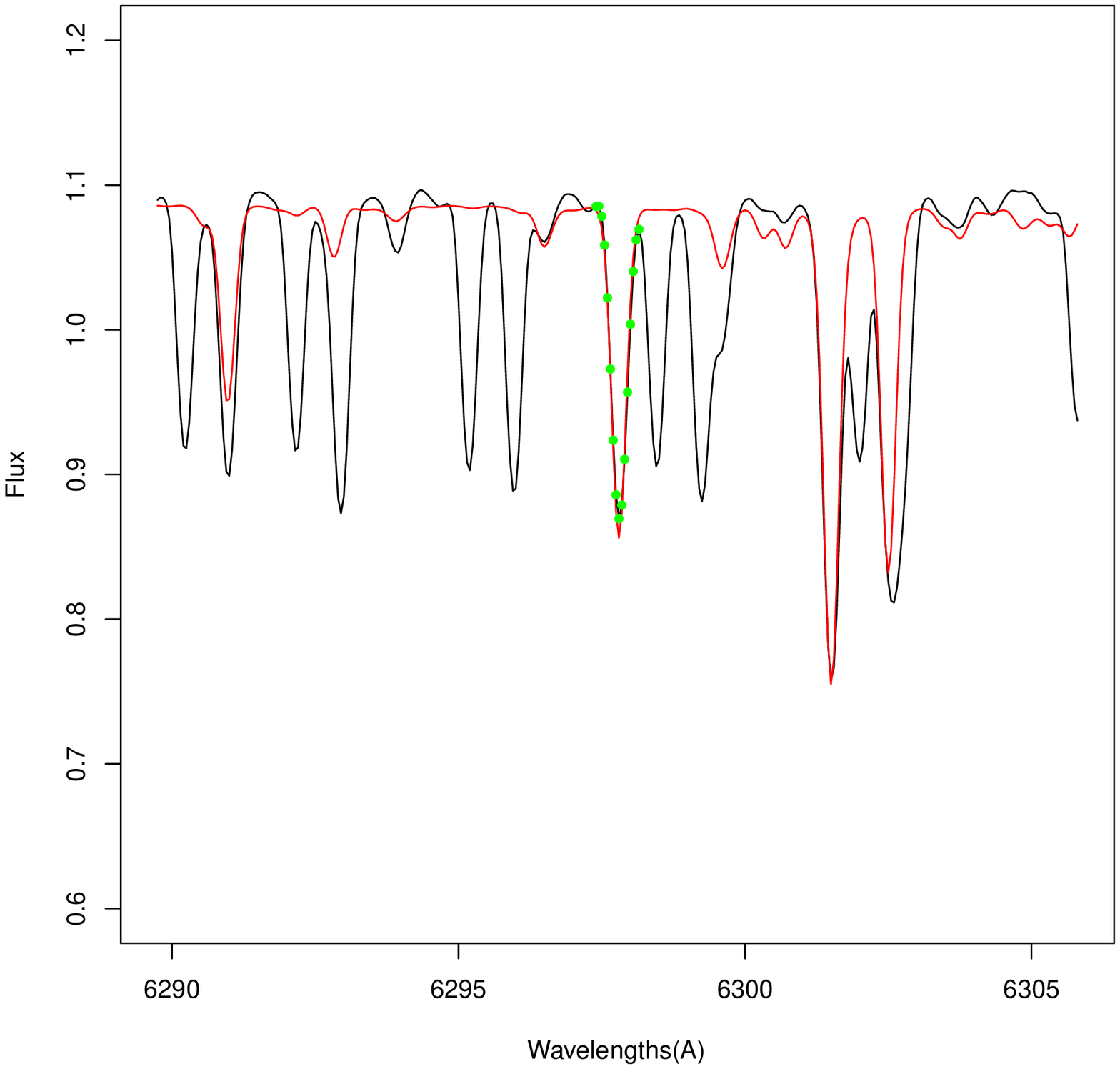}
\caption{Two examples: (i) pixel identification in the reference line (green crosses),
(ii) pseudo-continuum selection (blue circles), and (iii) vertical adjustment of the synthetic reference
spectrum (red line) to the studied spectrum (black line). In the top plot, the pseudo-continuum
area (blue circles) were selected automatically by SPADES and were used for the vertical
adjustment. In the bottom plot, no valid pseudo-continuum area was found.
The pixels in the line (green crosses) were used for the vertical adjustment.} 
\label{author1:fig5}
\end{figure}

\subsubsection{Line and continuum identification and vertical adjustment}
\label{author1:subsect1}
For each spectral line, the pixels constituting the line are dynamically
identified using the synthetic spectra. Using the rest wavelength of the line,
SPADES starts from its center and agregates pixels in each wing as long as the
value of the signal increases. If less than five pixels are associated to
the line, it is considered too weak and rejected from the subsequent analysis.
This parameter is adjustable by the user. Figure~\ref{author1:fig5} shows
two examples of sets of pixels associated to a line, i.e. green crosses on the figures.\\

SPADES relies on profiles fitting. This requires adjusting the observed and 
synthetic spectra on the same vertical scale. It is possible, in the pre-processing,
to normalize the overall continuum to 1. This is a global adjustment on the whole
spectrum. This is not enough for the comparison of two lines. Therefore, SPADES
systematically adjusts locally, line by line, the vertical level of the synthetic
line to the one of the observed line.

SPADES first identifies flat areas (pseudo-continuum) in the synthetic spectrum in the vicinity of the
studied line and selects one on each side of the line. It then checks that these two
areas also correspond to pseudo-continuum in the studied spectrum. If the check is positive,
 it uses these areas to vertically scale the synthetic spectrum on the studied
one, by linear regression, i.e. there are two degrees of freedom, vertical scaling, and
slope.

If there are no flat areas in the synthetic spectrum or the studied spectrum contains (a) line(s) in the selected pseudo-continuum, another
method is used for the vertical adjustement. The method depends on the diagnostic.
For all diagnostics relying on the quadratic sum of the residuals (e.g. the iron abundance,
~Sect.~\ref{ironAbundance}), the pixels in the lines are 
directly used for the vertical adjustment (allowing here for a single
degree of freedom, the vertical scale). For the diagnostic that relies on the
the sum of residuals (i.e. the surface gravity, ~Sect.~\ref{SurfaceGravity}),
SPADES uses pre-tabulated pseudo-continuum.

An exception to this scheme is the $H_\alpha$ line used to derive
of the effective temperature, for which the pseudo-continuum and line
area are pre-tabulated.

Figure~\ref{author1:fig5} shows the local vertical adjustement of a synthetic
spectrum (red) to a studied spectrum (black). In the top plot, the pseudo-continuum
area (blue circles) were selected automatically by SPADES and were used for the vertical
adjustment of the synthetic reference spectrum (red line) to the studied spectrum
(black line). In the bottom plot, no valid pseudo-continuum area was found.
The pixels in the line (green crosses) were used for the vertical adjustment.

\begin{figure}[ht!]
\includegraphics[width = 270 pt]{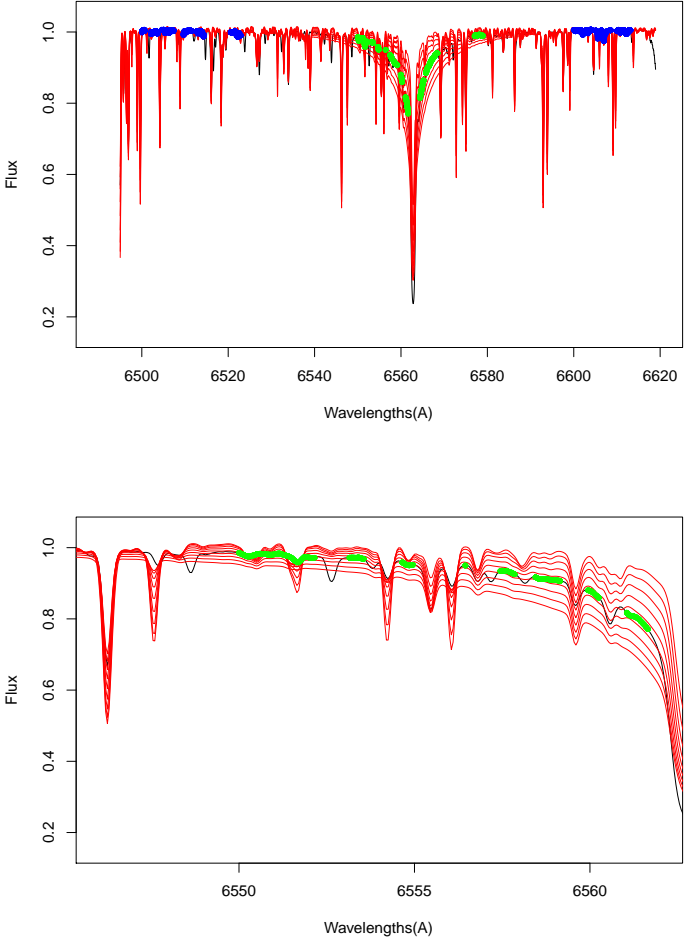}
\caption{Comparison of the $H_\alpha$ wings in the studied spectrum (black line) with the
1D grid of synthetic spectra (red lines) ranging here from 5000 (narrowest wings)
to 6600 K (broadest wing). The pseudo-continuum area is labeled with blue dots and thehave been
valid pixels in the $H_\alpha$ wings are identified with green crosses. The bottom plot
is a zoom on the blue wing.} 
\label{author1:fig7}
\end{figure}

\subsubsection{Effective temperature}
\label{EffectiveTemperature}
In this article, the stars considered are mid-F-G dwarfs, the wings of the Balmer lines are very sensitive
to the effective temperature \citep{VantveerMegessier1996, Fuhrmann1998, Fuhrmann2004, Fuhrmann2008, BarklemEtAl2002, Cayrel11}. In this study, SPADES used the wings of $H_\alpha$ present
in Giraffe setup HR14B to derive the effective temperature. The core of $H_\alpha$ was excluded because it is less sensitive
to \Teff variation and poorly modeled, specially using 1D LTE models, which is the case here.

As presented in section~\ref{Overview}, a 1D grid of synthetic spectra was calculated,
differing only by the effective temperature. After pre-processing
and vertical adjustment, each synthetic spectrum was compared to the studied
spectrum. For each synthetic spectrum, the degree of similarity of the synthetic
$H_\alpha$ wings with the studied one, is calculated as the quadratic sum
of the residuals between the pixels in the two lines:

\begin{equation}
\label{res2}
Sim_{H_\alpha} =  {\sum^{N}_{i = 0} (S(i)-R(i))^{2}},
\end{equation}
where $N$ is the number of valid pixels in the $H_\alpha$ line, $S(i)$ (resp. $R(i)$ the counts of the studied spectrum
(resp. reference spectrum) for pixel i.  

Figure \ref{author1:fig7} shows the comparison of the $H_\alpha$ wings in the studied spectrum (black line) with the
1D grid of synthetic spectra (red lines) ranging here from 5000 (narrowest wings)
to 6600 K (broadest wing). The pseudo-continuum area is labeled with blue dots and the
valid pixels in the $H_\alpha$ wings are identified with green crosses. The bottom plot
is a zoom on the blue wing.\\

\begin{figure}[ht!]
\includegraphics[width = 270 pt]{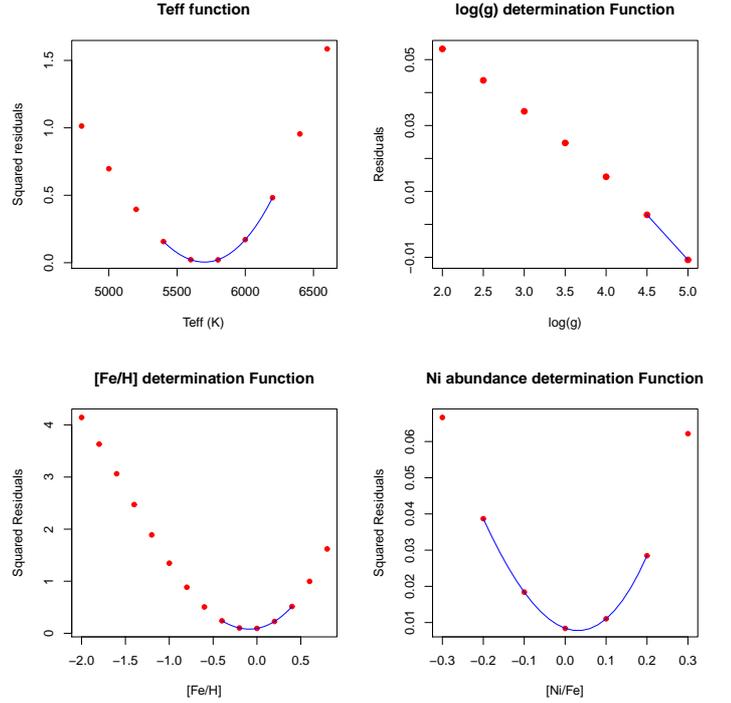}
\caption{Examples, for a synthetic studied spectrum, of similarity functions as a
function of effective temperature (top left), surface gravity (top right), iron over
hydrogen ratio (bottom left) and nickel over iron ratio (bottom right). The polynomial
(or linear interpolation for the surface gravity) used to minimize the functions are
represented as blue lines.} 
\label{author1:fig8}
\end{figure}

\begin{figure}[ht!]
\includegraphics[width = 270 pt]{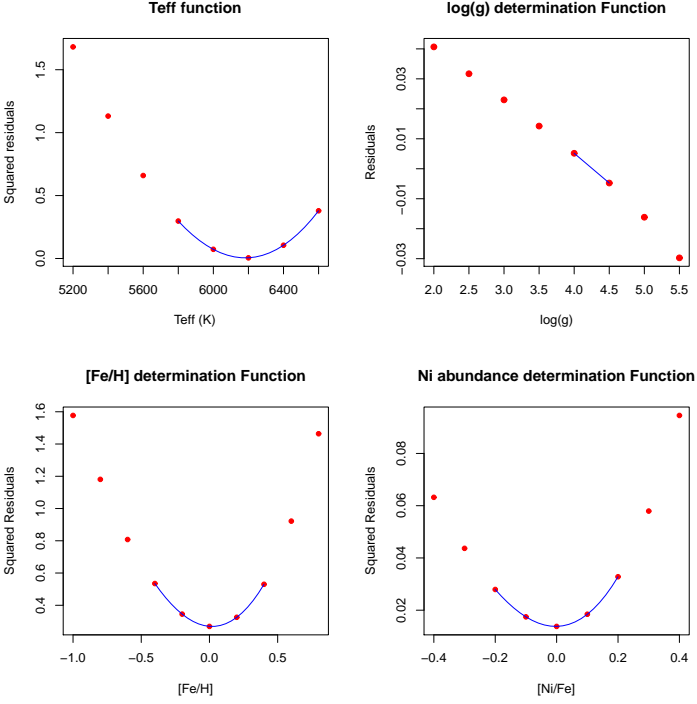}
\caption{ $\nu$~And similarity functions as a function effective temperature (top left),
surface gravity (top right), iron over hydrogen ratio (bottom left), and nickel over iron
ratio (bottom right). The polynomial (or linear interpolation for the surface gravity)
used to minimize the functions are represented as blue lines.} 
\label{similarityNuAnd}
\end{figure}

The degrees of similarity versus the effective temperature define a function
that is minimum for the effective temperature of the studied source. The minimum
is found by fitting a second-degree polynomial around the lowest similarity value. 
The top left plots in Figure~\ref{author1:fig8} and \ref{similarityNuAnd} give
examples of similarity functions as a function of effective temperature,
for a synthetic spectrum and for the ground-based observed spectrum of $\nu$~And respectively.
Each synthetic
spectrum in the 1D subgrid corresponds to a red dot. The second-degree
polynomial fitted to the similarity function to minimize it is represented as a blue line.\\

The similarity functions for the synthetic spectra and $\nu$~And show 
similar degrees of regularity and smoothness. The first reason is that the
spectrum of $\nu$~And has a good S/N ratio. The second reason
is that in each point (red dots) the similarity functions derives from the 
comparison of the same noisy studied spectrum with different noiseless
synthetic spectra. The noises on the points of a similarity function are 
correlated. The noise in the studied spectrum and the systematic differences
between the studied and synthetic spectra propagate to the similarity
functions, not so much as fluctuation and irregularities in the functions,
but as a vertical and horizontal shift and as a flattening of the curvature
of the functions.

\subsubsection{Surface gravity by ionization equilibrium}
\label{SurfaceGravity}
The ionization equilibrium method is based on the fact that the abundances determined
from lines of the same element, but with different ionization stages, should be the same.	
Building on this idea, SPADES simultaneously models Fe lines in two ionization stages:
FeI and FeII.

Similarly as for \Teff, a 1D grid of synthetic spectra, differing only by the surface gravity
(this time) was calculated. For each a ``degree of ionization equilibrium'' (hereafter refered to as ``gravity
similarity'') is calculated as

\begin{equation}
Sim_{\mathrm grav} =  \frac{1}{N_{FeI}}{\sum^{N_{FeI}}_{l = 0} Res_{l}} - \frac{1}{N_{FeII}} {\sum^{N_{FeII}}_{l = 0} Res_{l}},
\end{equation}
where $N_{FeI}$ (resp. $N_{FeII}$) are the numbers of FeI (resp. FeII) lines used and $R_{i}$ the count residuals for the
line l:
\begin{equation}
\label{res2}
Res_{l} =  {\sum^{N}_{i = 0} (S(i)-R(i))},
\end{equation}
where $N$ is the number of valid pixels in the line, $S(i)$ (resp. $R(i)$ the count of the studied spectrum 
(resp. reference spectrum) for pixel i.

\begin{figure}[ht!]
\includegraphics[width = 250 pt]{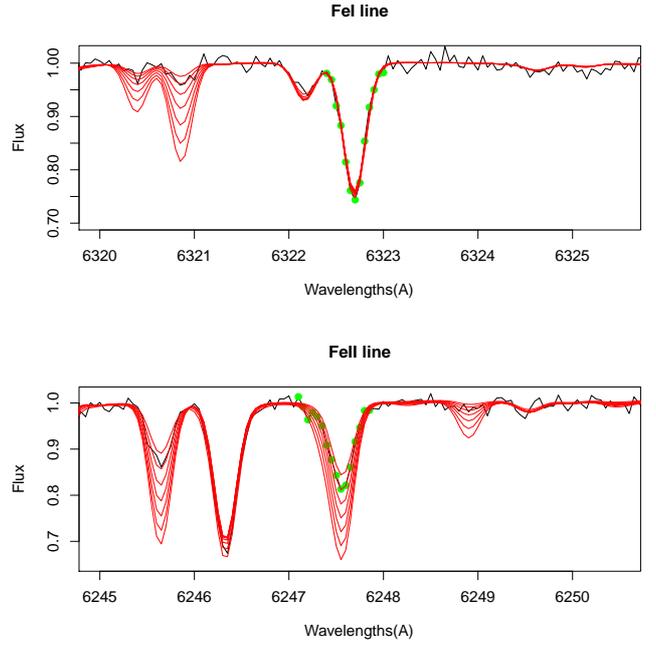}
\caption{FeI line (top, green crosses) and an FeII line (bottom, green crosses),
used by SPADES to determine of \logg. The studied spectrum (\logg = 4.5) is given in black and the
red lines are the synthetic reference spectra of the 1D subgrid ranging from \logg = 2.0 to 5.0
by step of 0.5.} 
\label{loggLines}
\end{figure}

Figure \ref{loggLines} shows an FeI line (top, green crosses) and an FeII line (bottom, green crosses),
used by SPADES to determine of \logg. The studied spectrum (\logg = 4.5) is denoted in black and the
red lines are the synthetic reference spectra of the 1D subgrid ranging from \logg = 2.0 to \logg = 5.0
by steps of 0.5. The neutral iron line is alsmost insensitive to gravity, while the singly ionized iron
line becomes stronger for decreasing gravities.

The gravity similarity, as defined above, should be null for the surface gravity of the studied source.
The zero of the gravity similarity function is found by interpolation. The top right plots in
Figure~\ref{author1:fig8} and figure~\ref{similarityNuAnd} give an example of similarity functions as a function of surface gravity.
Each synthetic spectrum in the 1D subgrid corresponds to a red dot. The linear interpolation
used to find the gravity corresponding to the zero gravity similarity is represented as a blue line.

\subsubsection{iron over hydrogen ratio}
\label{ironAbundance}
The \FeH determination is based on a global FeI line profile fitting. For the other parameters,
a 1D grid of synthetic spectra, differing only by the iron abundance, was calculated.
For each synthetic spectrum, the global similarity between all synthetic neutral iron lines and
all observed neutral iron lines is derived as

\begin{equation}
Sim_{Fe} =  {\sum^{N_{FeI}}_{l = 0} Res^2_{l}}, 
\end{equation}
where $N_{FeI}$ is the number of neutral iron lines in the list of reference lines,
$Res^2_{l}$ is the square of the count residuals for line l, given by formula~\ref{res2}. 

The iron content of the source corresponds to the minimum similarity degree. It is found by adjusting
a third-degree polynomial through the ``iron similarity function''. A third-degree polynomial is
preferred over a second-degree polynomial, because the ``iron similarity function'' is usually asymmetric:
steeper on the metal-rich side than on the metal-poor side. The bottom left plots in Figure \ref{author1:fig8} and figure~\ref{similarityNuAnd}
give an example of similarity functions as a function of the iron over hydrogen ratio. Each synthetic spectrum
in the 1D subgrid corresponds to a red dot. The third-degree polynomial fitted to the similarity function to
minimize it is represented as a blue line.

\subsection{Third step: Individual abundances}
\label{IndividualAbundances}
Once the iterative determination of the three atmospheric parameters has converged,
the individual abundances are determined. The general principle is very similar to
the determination of the iron content, except that the determination is direct and
not within an iterative loop with several other parameters.

Each element was determined individually, using a different pre-defined list of lines.
The difference with the \FeH determination method is that for each element, the 1D
grid was calculated on-the-fly for the \Teff, \logg, and \FeH determined by SPADES and varying
only by the abundance of the element to be determined. SPADES interpolates on-the-fly the
input files for the Kurucz programs (e.g. the opacity distribution functions - ODF
\citep{CastelliKurucz2004}) and calls Kurucz ATLAS9 \citep{Kurucz2005}
to calculate the atmospheric models and Kurucz SYNTHE \citep{Kurucz2005}
to compute the synthetic spectra. The GNU-Linux ported versions of ATLAS9 and SYNTHE were used
\citep{SbordoneEtAl2004, SbordoneEtAl2005}.

Element per element, for each synthetic spectrum the global similarity between all synthetic
lines and all observed lines of the element X is derived as

\begin{equation}
Sim_{X} =  {\sum^{N_{X}}_{l = 0} Res^2_{l}}, 
\end{equation}
where $N_{X}$ is the number of lines of the element X in the list of reference lines and
$Res^2_{l}$ is the square of the count residuals for line l, given by formula~\ref{res2}. 

The abundance of the element X corresponds to the minimal similarity degree. It is found by adjusting
a third-degree polynomial through the element X similarity function. The bottom right plots in
Figure \ref{author1:fig8} and figure~\ref{similarityNuAnd} give an example of similarity functions as a function of nickel over iron ratio.
Each synthetic spectrum in the 1D subgrid corresponds to a red dot. The third-degree
polynomial fitted to the similarity function to minimize it is represented as a blue line.

\section{Auxiliary data}
\label{author1:sect2}

\subsection{Reference grid of models and synthetic spectra}
\label{referenceGrid}
The unidimensional reference grids used to derive the stellar parameters
are calculated by SPADES by interpolation using a pre-calculated grid.
Models and synthetic spectra were calculated using the Kurucz programs
ATLAS9 and SYNTHE \citep{Kurucz2005} ported to GNU-Linux \citep{SbordoneEtAl2004, SbordoneEtAl2005}.
About 2000 models and spectra were generated over the following range of parameters:
\begin{itemize}
\item Effective temperature: \Teff from 4800 to 6800 K with a step of 200 K. \\
\item Surface gravity: \logg from 2.0  to 5.5 with a step of 0.5.\\
\item iron abundance: \FeH from -3.0 to 1.0 dex with a step of 0.20 dex.\\
\item Microturbulence: $\xi$ = 1 \kms
\end{itemize}

The models were calculated with the overshooting option switched off and $l/H_p = 1.25$.
The convergence of each model was checked. The spectra were calculated for the spectral
domain ranging from 600 to 680 nm, corresponding to the Giraffe/VLT HR13 and HR14 setups
and a resolving power R~=~300~000 (SPADES convolves and re-samples the synthetic
spectra to the resolving power(s) of the studied spectrum).

\subsection{List of reference lines}
The reference lines used to derive the surface gravity, iron abundance, and individual
abundances were extracted from the line list of \citet{Bensby03}. We note
that the astrophysical oscillator strengths provided by \citet{Bensby03} were not used
in calculating of the synthetic spectra (see~Sect.~\ref{referenceGrid}). For
consistency, we used the Kurucz oscillator strengths for all lines (reference or not).


\section{Performance assessment}
\label{author1:sect3}
To assess the performances of SPADES, three series of tests were performed.\\

The aim of the first series was to assess the internal systematic and random
errors obtained as a function of the S/N ratio, but independently of
the question of the initialization parameters. We also investigated the
correlations between the errors on the different parameters.

In the second tests, high S/N ground-based observed spectra were
analyzed with SPADES to assess the external errors.

SPADES proceeds iteratively to derive the atmospheric parameters. The
third series of tests aims to assess the convergence properties of SPADES
when it is initialized with parameters that are significantly off the true
values.\\

All tests were performed for the wavelength range and the resolving power
of the HR13 (R~=~22~500) and HR14B (R~=~28~000) Giraffe setups. The three series
of tests are presented below.

\begin{figure}[ht!]
\includegraphics[width = 270 pt]{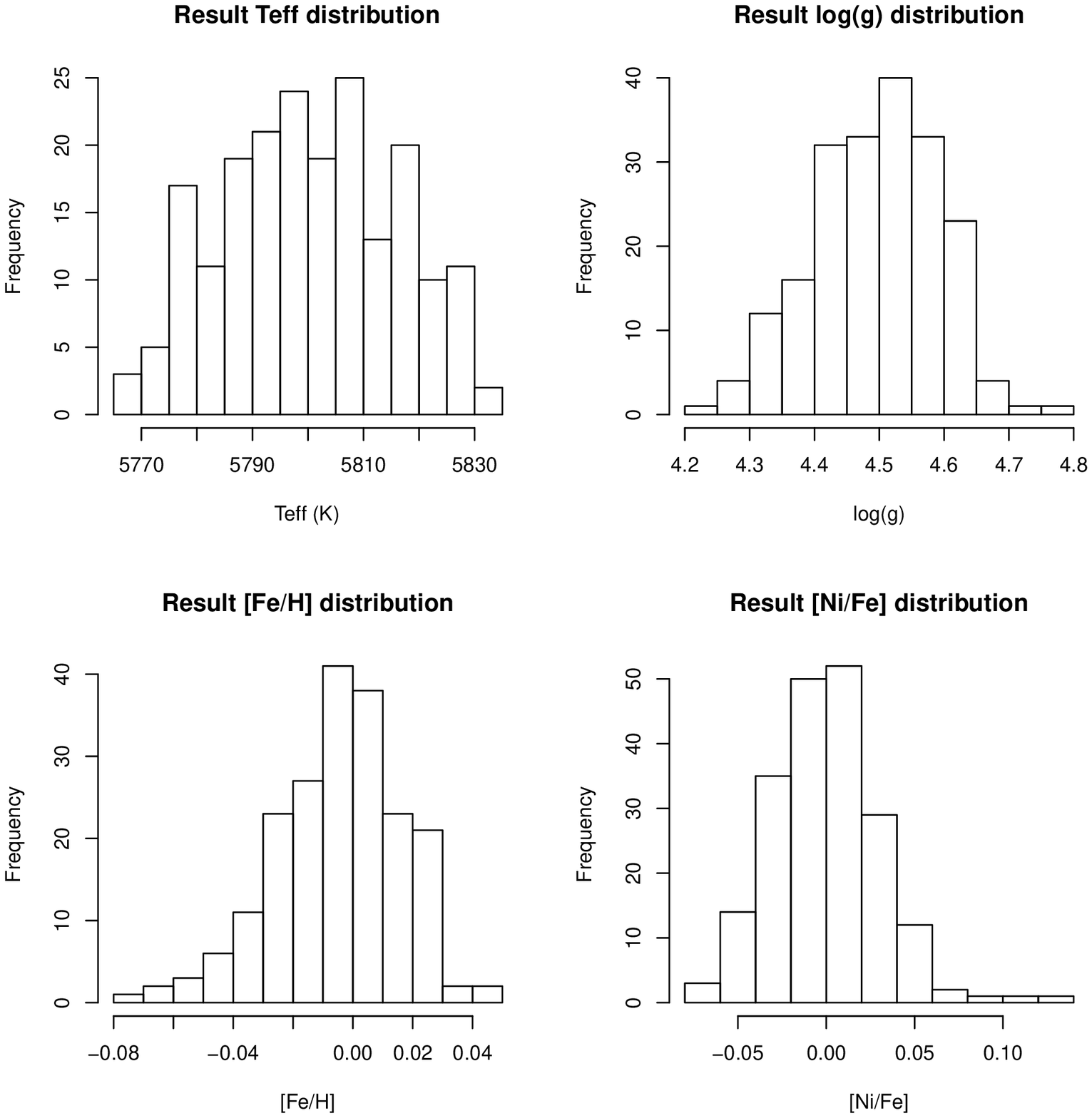}
\caption{Distributions of the \Teff (top left), \logg (top right), \FeH (bottom left)
and \NiFe (bottom right), estimated by SPADES for 200 S/N=50 simulated HR13 and HR14B Giraffe
spectra. The true parameters are \Teff~=~5800~K, \logg~=~4.5, \FeH~=~0.0~dex, and \NiFe~=~0.0~dex} 
\label{author1:hists}
\end{figure}

\subsection{Internal errors}
\label{InternalErrors}
The internal systematic and random errors were assessed by Monte Carlo method. Starting from a synthetic
spectrum from the reference grid with \Teff~=~5800~K, \logg~=~4.5, \FeH~=~0.0~dex, $\xi$~=~1~\kms
and solar abundance scale, three series of 200 noisy spectra were generated for SNRs 30, 50 and
100. The noise was injected following a Poisson distribution. The 600 spectra were analyzed
with SPADES to estimate their \Teff, \logg, \FeH as well as \TiFe and \NiFe ratios as examples
of abundance ratios estimates. In this first test, SPADES was initialized with the
true parameters of the synthetic spectra. Table~\ref{table:2} presents the medians and dispersions
of the distributions of residuals, i.e. estimated minus true, of the parameters derived by SPADES
for each S/N ratio. Figure~\ref{author1:hists} shows the distributions of \Teff,
\logg, \FeH, and \NiFe estimated by SPADES for the 200 S/N=50 spectra.\\

The distributions of temperature and gravity residuals present no significant biases. The systematic
errors on the iron over hydrogen and titanium and nickel over iron ratios are of about a few
milli-dex and are most of the time one order of magnitude smaller than the random errors. The random
errors are small, e.g. at S/N=30: less than 30~K for \Teff, 0.15 for \logg and a few hundredths of
dex for the abundance ratios. we emphasize that these are internal errors accounting only
for photon noise. External errors are assessed in Sect.~\ref{ExternalErrors}.

\begin{table}[ht!]
\caption{Medians and dispersions of the distributions of residuals, i.e. estimated minus true,
on \Teff, \logg, \FeH, \TiFe and \NiFe derived by SPADES for the S/N: 100,
50 and 30.}
\label{table:2} 
\begin{tabular}{l l | r r r}
                       &       & S/N=100 & S/N=50 & S/N=30 \\ \hline
med(\Teff$_{res}$)     & (K)   &  0      &  0     &  0     \\
$\sigma$ \Teff$_{res}$ & (K)   &  8      &  16    &  27    \\ \hline
med(\logg$_{res}$)     &       &  0.00   &  0.00  &  0.00  \\
$\sigma$ \logg$_{res}$ &       &  0.05   &  0.09  &  0.15  \\ \hline
med(\FeH$_{res}$)      & (dex) &  0.002  & -0.001 & -0.003 \\
$\sigma$ \FeH$_{res}$  & (dex) &  0.009  &  0.020 &  0.030 \\ \hline
med(\TiFe$_{res}$)     & (dex) &  0.004  &  0.005 &  0.005 \\
$\sigma$ \TiFe$_{res}$ & (dex) &  0.003  &  0.040 &  0.060 \\ \hline
med(\NiFe$_{res}$)     & (dex) & -0.004  &  0.002 &  0.010 \\
$\sigma$ \NiFe$_{res}$ & (dex) &  0.010  &  0.030 &  0.050 \\
\end{tabular}
\end{table}

\begin{figure}[ht!]
\includegraphics[width = 270 pt]{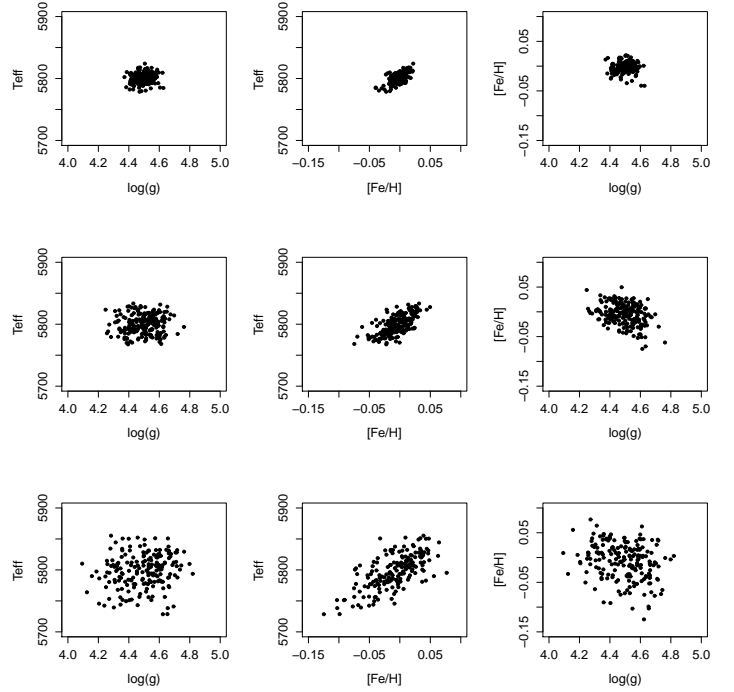}
\caption{SPADES estimate of \Teff, \logg, and \FeH versus each others for the 200 stars with S/N=100 (top),
S/N=50 (middle), and S/N=30 (bottom).} 
\label{author1:corr}
\end{figure}

Figure~\ref{author1:corr} presents the estimate by SPADES of \Teff, \logg, and \FeH versus
each other for the 200 stars with S/N=100 (top), S/N=50 (middle), and S/N=30 (bottom). The clearer error
correlation is between the effective temperature and the iron abundance: an error of $-$50~K
corresponds to an error of about $-$0.075~dex. This correlation, as the others, is intrinsic
to the physics of the stellar atmopsheres. While $H_\alpha$ is weakly sensitive to metallicity
\citep{VantveerMegessier1996}, the equivalent width of the iron lines is sensitive
to the effective temperature.

\subsection{External errors: ground-based observed spectra}
\label{ExternalErrors}
The first series of tests was devoted to internal errors. The source of noise was Poissonian photon
noise. The aim of the second series was to assess the external systematic errors. There are several
possible sources of external systematics. In particular, the physics of the stellar atmospheres do
not reproduce the reality perfectly (e.g. discrepant oscillator strengths, local thermodynamical
Equilibrium assumption, 1D modeling of the atmosphere), there are usually instrumental
calibration residuals (e.g. from bias, flat-field, possibly fringing, etc.) and residuals from the
observing conditions (e.g. from sky-subtraction).\\

\subsubsection{The Giraffe solar spectrum}
To investigate these problems, we retrieved from the Giraffe archive \citep{RoyerEtAl2005, RoyerEtAl2008}
the high S/N ratio HR13 (R~=~22~500) and HR14B (R~=28~000) observed solar spectrum. The
Giraffe solar spectrum was acquired in twilight conditions, which could alter the equivalent
widths \citep{MolaroEtAl2008}. Moreover, the library of reference spectra used by SPADES was calulated with the Kurucz
oscillator strengths ($\log$~gf), not with astrophysical $\log$~gf, even for the reference lines used
by the different diagnostics. We therefore proceeded in two steps to analyze the Giraffe solar spectrum.
First the effective temperature and surface gravity were fixed to their bibliographic values, i.e.
$\Teff = 5777$ K and $\logg = 4.44$, and only the iron over hydrogen and titanium and nickel over iron ratios were determined
by SPADES. Then, in a second step, the solar spectrum was re-analyzed, this time requesting SPADES to
determine the five parameters \Teff, \logg, \FeH, \TiFe, and \NiFe.

Table~\ref{table:3} lists the iron over hydrogen and titanium and nickel over iron ratios derived by SPADES
when the effective temperature and surface gravity were fixed to their bibliographic values
$\Teff = 5777$ K and $\logg = 4.44$. The iron over hydrogen and titanium over iron ratios are underestimated
by $-$0.08 and $-$0.09~dex respectively, while the nickel over iron ratio is close to the true value.

\begin{table}[ht!]
\caption{iron over hydrogen and titanium and nickel over iron ratios derived by SPADES for a solar HR13
and HR14B solar spectrum, when the effective temperature and surface gravity were fixed to their
bibliographic values $\Teff = 5777$ K and $\logg = 4.44$.}             
\label{table:3}
\centering       
\begin{tabular}{c c c}
 \FeH  & \TiFe  & \NiFe \\
 (dex) & (dex)  & (dex) \\ \hline	               
 -0.08 & -0.09  & -0.03 \\      
\end{tabular}	
\end{table}

The solar spectrum was then re-analyzed, requesting SPADES to also determine \Teff and \logg.
SPADES was initialized with the bibliographic parameters Sun and converged in five iterations.
Table~\ref{table:4} presents the parameters obtained by SPADES for the Giraffe solar spectrum.

\begin{table}[h!]
\caption{Effective temperature, surface gravity, iron over hydrogen and titanium and nickel over iron ratios
derived by SPADES for the Giraffe HR13 and HR14B solar spectra.}             
\label{table:4}
\centering       
\begin{tabular}{c c c c c}
\Teff & \logg &  \FeH  & \TiFe & \NiFe \\
 (K)  &       &	(dex)  & (dex) & (dex) \\ \hline            
5689  &  4.46 & -0.18  & -0.06 & 0.04  \\          
\end{tabular}	
\end{table}

\begin{figure}[ht!]
\includegraphics[width = 270 pt]{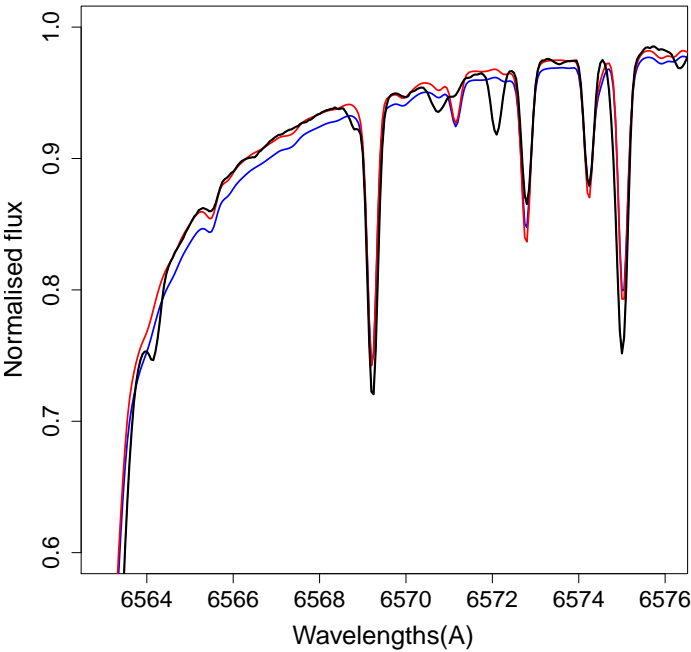}
\caption{Zoom on the red wing of the $H_\alpha$ line from the Giraffe HR14B solar spectrum (black line).
Two synthetic spectra are overplotted: in red a synthetic spectrum with \Teff~=~5689~K and in blue
\Teff~=~5777~K (the values of the other parameters are those of the Sun: \logg~=~4.44 and \FeH~=~0.0~dex).} 
\label{author1:HalphaSun}
\end{figure}

The effective temperature derived by SPADES, 5689~K, is 88~K lower than the bibliographic temperature
\Teff~=~5777~K. Figure~\ref{author1:HalphaSun} shows a zoom on the red wing of the $H_\alpha$ line from the
Giraffe HR14B solar spectrum (black line). Two synthetic spectra are overplotted: in red a synthetic
spectrum with \Teff~=~5689~K and in blue \Teff~=~5777~K (the values of the other parameters are those
of the Sun: \logg~=~4.44 and \FeH~=~0.0~dex). The red spectrum (\Teff~=~5689~K) agrees much better
agreement with the Giraffe solar spectrum than the blue spectrum (\Teff~=~5777~K).

The surface gravity derived by SPADES, \logg~=~4.46 is consistent with the Sun's gravity \logg~=~4.44.

The iron to hydrogen ratio derived by SPADES is underestimated by $-$0.18~dex with respect to the Sun.
There are two main origins for this offset. On the one hand, there is the offset of $-$0.08~dex reported in
table~\ref{table:3} that is likely due to the modeling of the spectra, in particular slightly discrepant
oscillator strengths,
and also to possible equivalent width alterations due to twilight observing, as reported by \citet{MolaroEtAl2008}.
On the other hand, as discussed in Sect.~\ref{InternalErrors}, the error on the 
effective temperature propagates into an error on the iron abundance. The offset of $-$88~K here
propagates into an offset of $-$0.1~dex.

The titanium and nickel over iron ratios derived when \Teff and \logg are determined by SPADES are
similar to those obtained when \Teff and \logg were fixed to their bibliographic values. The reason is
that the iron, titanium and nickel abundances are all similarly affected by the $-$88~K offset and therefore
the effect partly canceled out in the logarithmic ratios of titanium over iron and nickel over iron.\\

\subsubsection{S4N reference spectra}
 The parameters of the Sun are very accurately known, which makes it a natural
target to assess the performances of SPADES. On the other hand, \citet{MolaroEtAl2008}
reported an alteration of the equivalent widths of the lines in the Giraffe solar spectrum,
which was acquired in twilight conditions.\\

To further test the performances of SPADES with a ground-based observed spectrum, we selected
5~stars, the Sun, $\nu$~And (HD9826), $\beta$~Vir (HD102870), $\mu$~Her (HD161797)
and $\sigma$~Dra (HD185144) in~\citet{Cayrel11}. The authors derived the
effective temperatures of their stars from angular diameters measured with interferometry.
All their stars have diameters known to 2\% or better, leading to an accuracy on
the effective temperatures of about 25~K. All stars are bright and were
extensively studied. \citet{Cayrel11} adopted for \logg and \FeH recent determinations
from the PASTEL catalog \citep{Soubiran10}. We proceeded similarly, adopting for
\logg and \FeH the median over the five most recent determinations contained in PASTEL.
The titanium and nickel over iron ratios were extracted from \citet{ValentiFischer2005}.
The spectra for the five stars were retrieved from the S4N library \citep{AllendePrietoEtAl2004}.
They were degraded in resolving power and in sampling and restricted in wavelength
to the Giraffe HR13 and HR14B setups.\\

\begin{figure}[ht!]
\includegraphics[width = 270 pt]{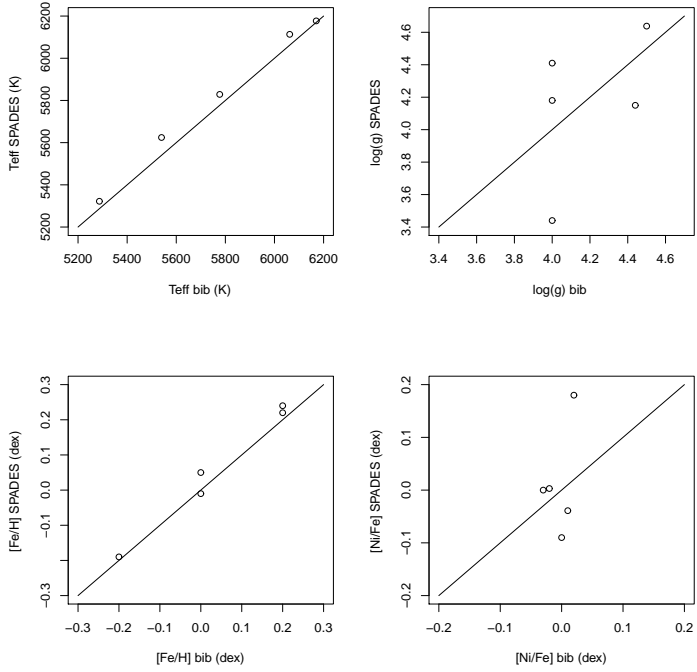}
\caption{Temperatures (top left), gravities (top right), iron over hydrogen
(bottom left), and nickel over iron ratios (bottom right) estimated by SPADES
versus their bibliographic values.} 
\label{S4Nplot}
\end{figure}

SPADES converged in two ($\nu$~And) to four ($\mu$~Her and $\sigma$~Dra) iterations.
Table~\ref{S4Nresults} lists the bibliographic parameters and the parameters
derived by SPADES for the five tests stars. Fig.~\ref{S4Nplot} displays the 
temperatures, gravities, iron over hydrogen and nickel over iron ratios
estimated by SPADES versus their bibliographic values. Table~\ref{S4Nresiduals}
presents the mean and dispersion of the residuals (SPADES minus bibliographic)
on the estimation of the atmospheric parameters as well as the titanium and nickel
over iron ratios.

\begin{table}[h!]
\caption{ Bibliographic parameters and parameters derived by SPADES
for the five tests stars.}             
\label{S4Nresults}
\centering       
\begin{tabular}{l c c c c c}
name                    & \Teff & \logg &  \FeH  & \TiFe & \NiFe      \\
                        & (K)   &       & (dex)   & (dex)   & (dex)   \\ \hline
Sun$_{bib}$             & 5777  & 4.44  &    0.00 &    0.00 &    0.00 \\
Sun$_{SPADES}$          & 5829  & 4.15  &    0.05 & $-$0.12 & $-$0.09 \\ \hline
$\nu$ And$_{bib}$       & 6170  & 4.12  &    0.06 &    0.02 & $-$0.02 \\
$\nu$ And$_{SPADES}$    & 6177  & 4.18  & $-$0.01 & $-$0.11 &    0.00 \\ \hline
$\beta$ Vir$_{bib}$     & 6062  & 4.11  &    0.16 & $-$0.01 &    0.01 \\
$\beta$ Vir$_{SPADES}$  & 6113  & 3.44  &    0.22 & $-$0.18 & $-$0.04 \\ \hline
$\mu$ Her$_{bib}$       & 5540  & 3.99  &    0.28 & $-$0.08 &    0.02 \\
$\mu$ Her$_{SPADES}$    & 5624  & 4.41  &    0.24 &    0.06 &    0.18 \\ \hline
$\sigma$ Dra$_{bib}$    & 5287  & 4.57  & $-$0.23 &    0.00 & $-$0.03 \\
$\sigma$ Dra$_{SPADES}$ & 5322  & 4.64  & $-$0.19 &    0.05 &    0.00 \\ \hline
\end{tabular}	
\end{table}

\begin{table}[h!]
\caption{Means and dispersions of the residuals (SPADES minus bibliographic)
on the estimation of the atmospheric parameters as well as the titanium and nickel
over iron ratios.}           
\label{S4Nresiduals}
\centering
\begin{tabular}{l c c c c c}
         & \Teff & \logg   & \FeH  & \TiFe & \NiFe  \\
         & (K)   &         & (dex) & (dex) & (dex)  \\ \hline
mean     & 46    & $-$0.08 & 0.01  & $-$0.05 & 0.01 \\
$\sigma$ & 25    &  0.37   & 0.05  &    0.12 & 0.09 \\
\end{tabular}
\end{table}

On average, the effective temperature is recovered with a bias of $+$46~K.
For the same stars, \citet{Cayrel11} found a bias of similar amplitude, but
with the  opposite sign, of about $-$90~K.
In addition to the small statistics and the details of the H$_\alpha$ profile
fiting, we note that the physics used to compute the grids
of models and synthetic spectra differ in several aspects. SPADES relies
on Kurucz ATLAS9 models computed with the mixing length parameter
l/H$_p = 1.25$ and SYNTHE spectra (see Sect.~\ref{referenceGrid}).
\citet{Cayrel11} used the Kurucz ATLAS9 models with l/H$_p = 0.5$ and
a modified version of the Kurucz BALMER9 with the Stark broadening treatment
of \citet{StehleHutcheon1999} and impact broadening by neutral hydrogen
collisions of \citet{AllardEtAl2008}.

The iron over hydrogen and nickel over iron ratios are recovered without
significant offsets. The titanium over iron ratio, determined from four Ti lines
(without astrophysical log~gf adjustement), presents an offset of $-$0.05~dex.
To the limit of the small statistics, the error on \Teff shows no trend
with \Teff.\\

\subsection{Convergence properties}
\label{ConvergenceProperties}

\subsubsection{First test}
\label{random}
SPADES proceeds iteratively to derive the atmospheric parameters \Teff, \logg, and \FeH.
In the first series of tests, SPADES was initialized with the true parameters. Of
course, in reality, the true parameters are unknown. The aim of this third series
of tests is to assess the behavior of SPADES when it is initialized with parameters that
are significantly different from the true values.

\begin{figure}[ht!]
\includegraphics[width = 270 pt]{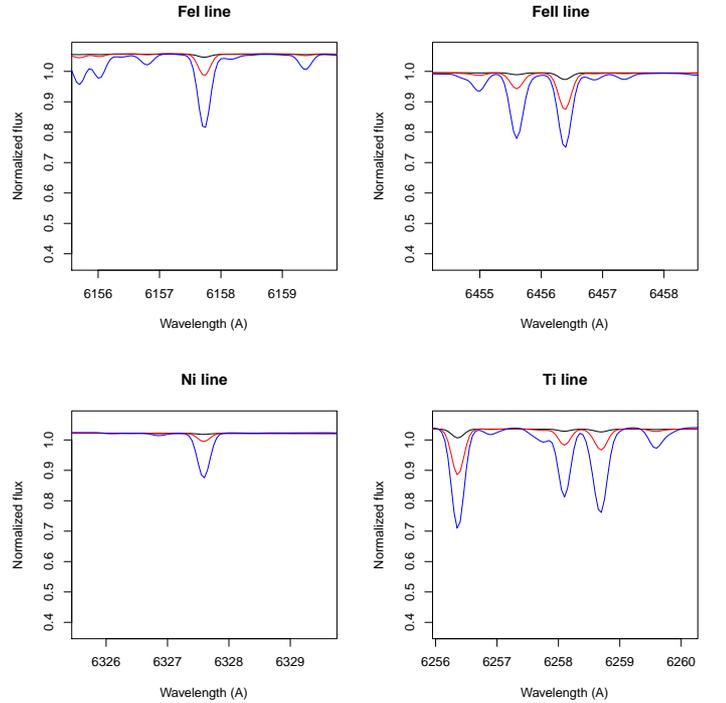}
\caption{Extract of three synthetic spectra of metallicity \FeH=0.0~dex (blue), \FeH=-1.0~dex (red), \FeH=-2.0~dex (black),
showing the influence of the metallicity on an Fe~I line (top left), an Fe~II line at 6456.39 \AA\ (top right), a Ni~I line (bottom left)
and a Ti~I line at 6258.11 \AA\ (bottom right).} 
\label{noLine}
\end{figure}

For this third test, 34 single stars were generated. Their parameters were randomly
drawn following uniform distributions over the following intervals: \Teff in $[5200, 6400]$~K,
\logg in [3.0, 4.5], and \FeH in [-2.0, 0.6]~dex. No star was simulated for \FeH below
-2.0, because the four Fe~II lines present in the Giraffe HR13 and HR14B become too weak
to allow derivating the surface gravity. This is illustrated in the top
right plot of Figure~\ref{noLine}, which shows three synthetic spectra of metallicity
$[Fe/H]=0.0$ dex (blue), $[Fe/H]=-1.0$ dex (red), $[Fe/H]=-2.0$ dex (black). The line at
6456.39 \AA\ is an Fe~II line. At \FeH~=~-2.0~dex, the line is barely visible.

\begin{figure}[ht!]
\includegraphics[width = 270 pt]{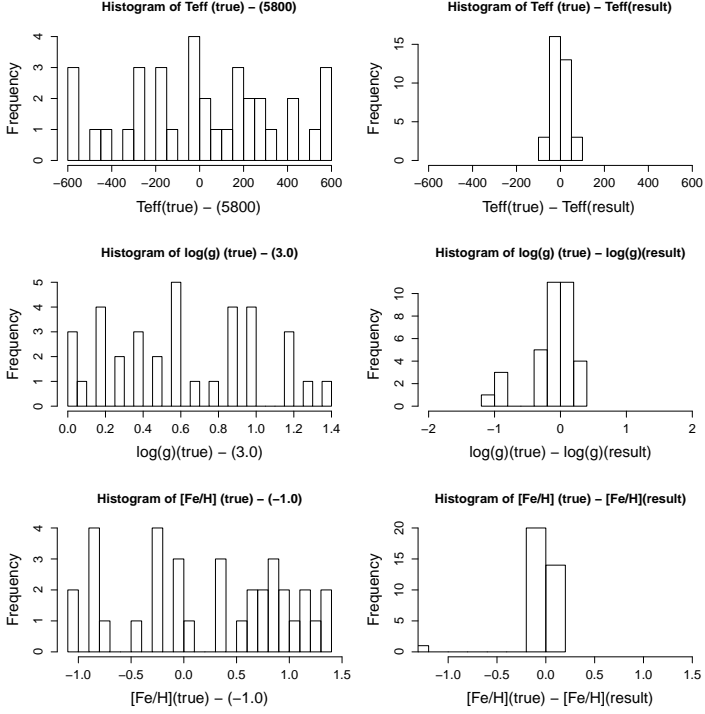}
\caption{Left: Differences between the true parameters of the 34 synthetic stars and the initialization
parameters: \Teff (top), \logg (middle) and \FeH (bottom). Right: Distributions of \Teff (top), \logg
(middle) and \FeH (bottom) residuals for S/N=50.} 
\label{HistTest3}
\end{figure}

For each of the 34 single stars, three noisy HR13 and HR14B Giraffe spectra were simulated for
S/N of 30, 50 and 100. The noise was Poissonian photon noise. The three series
of 34 spectra were analyzed by SPADES. In all 102 cases SPADES was initialized with the same
parameters: \Teff~=~5800~K, \logg~=~3.0 and \FeH~=~$-$1.0~dex. The left side of Figure~\ref{HistTest3}
presents the differences between the true parameters of the 34 synthetic stars and the initialization
parameters: \Teff (top), \logg (middle) and \FeH (bottom).

SPADES converged for 33 out of the 34 stars in at most six iterations (for all SNRs). It did not converge for one star with the parameters: \Teff~=~5386~K,
\logg~=~4.3, and \FeH~=~0.4~dex. At the first iteration, the difference of 500~K between the true \Teff and the initialization value
and the high true metallicity \FeH~=~0.4~dex, led SPADES to look for a synthetic spectrum with a metallicity higher than 1~dex, i.e.
beyond the upper boundary of the library of synthetic spectra. Unable to find the synthetic spectrum, SPADES stopped processing this
spectrum (at all three SNRs).

In all 33 converging stars, it was observed that the effective temperature converged faster than the other parameters.
This is because in mid-F-G dwarfs the $H_\alpha$ wings are hardly sensitive to wrong \logg and
\FeH \citep{VantveerMegessier1996}, while on the contrary \logg and \FeH are sensitive to wrong \Teff. This suggests a possible
future evolution for SPADES, which could at the first iteration determine only \Teff and then at the next iterations determine
all three atmospheric parameters. This would probably have prevented SPADES from looking for spectra outside the boundaries of
the library for the 34$^{th}$ spectrum.

Table~\ref{tabConv} presents the medians and the dispersions of the distributions of \Teff, \logg, and \FeH residuals,
i.e. estimated minus true, derived by SPADES for each S/N ratio. The right side of Figure~\ref{HistTest3} shows the
distributions of \Teff (top), \logg (middle), and \FeH (bottom) residuals for S/N=50. The performances derived from
test~3 (for of a wider variety of input atmospheric parameters) are similar to those obtained in test~1.
The convergence process has therefore a very moderate impact on SPADES final \Teff, \logg, and \FeH.

\begin{table}[h!]
\caption{Medians and dispersions of the distributions of \Teff, \logg, and \FeH residuals,
i.e. estimated minus true, derived by SPADES for the SNRs 100, 50 and 30.}
\label{tabConv}
\begin{tabular}{l l | r r r}
                       &       & S/N=100 & S/N=50 & S/N=30 \\ \hline
med(\Teff$_{res}$)     & (K)   &     8 &   10 &     0 \\
$\sigma$ \Teff$_{res}$ & (K)   &    17 &   24 &    27 \\ \hline
med(\logg$_{res}$)     &       & -0.03 & 0.00 &  0.02 \\
$\sigma$ \logg$_{res}$ &       &  0.08 & 0.15 &  0.25 \\ \hline
med(\FeH$_{res}$)      & (dex) &  0.00 & 0.00 & -0.01 \\
$\sigma$ \FeH$_{res}$  & (dex) &  0.02 & 0.03 &  0.03 \\ \hline
\end{tabular}
\end{table}

Out of the 33 converged stars, \TiFe and \NiFe ratios were derived for 21. Below  ametallicity of about $-$1.4~dex
(the precise value is a function of \Teff), the lines contained in the HR13 and HR14B become too weak to derive
the abundances. This is illustrated by the bottom left and bottom right plots in Figure~\ref{noLine}. At $[Fe/H]=-1$~dex
(red line), the Ni and Ti lines (at 6258.11 \AA) are already weak and at $[Fe/H]=-2$~dex they are not visible anymore.

Table~\ref{table:8} presents the medians and the dispersions of the distributions of \TiFe and \NiFe residuals,
i.e. estimated minus true, derived by SPADES for the SNRs 100, 50 and 30. The systematic and random errors derived
from test~3 are on average larger than those obtained from test~1. The main reason is very likely that in test~3,
\TiFe and \NiFe ratios are obtained for a wide range of metallicities down to $[Fe/H]\sim-1.4$~dex, while in test~1
a solar abundance star was considered.

\begin{table}[h!]
\caption{Medians and dispersions of the distributions of \TiFe and \NiFe residuals, i.e. estimated minus true,
derived by SPADES for the SNRs 100, 50 and 30.}
\label{table:8}
\begin{tabular}{l l | r r r}
                       &       & S/N=100 & S/N=50 & S/N=30 \\ \hline
med(\TiFe$_{res}$)     & (dex) &  0.008 & 0.03 & 0.06 \\
$\sigma$ \TiFe$_{res}$ & (dex) &  0.020 & 0.07 & 0.12 \\ \hline
med(\NiFe$_{res}$)     & (dex) & -0.007 & 0.02 & 0.05 \\
$\sigma$ \NiFe$_{res}$ & (dex) &  0.030 & 0.07 & 0.07 \\
\end{tabular}
\end{table}

\subsubsection{Second test}
 Iterative methods present the risk to converge toward a local
minimum, rather than the global minimum. To investigate this
risk in more detail, we selected one out of the 34 times 200 spectra analyzed in
Sect.~\ref{random} above, with S/N=50. The true parameters of the 
spectrum were \Teff~=~5992~K, \logg~=~4.4, \FeH~=~$-$0.3~dex,
\TiFe~=~0~dex and \NiFe~=~0~dex. SPADES analyzed the spectrum a first
time, starting from its true parameters and found \Teff~=~6025~K,
\logg~=~4.35, \FeH~=~$-$0.26~dex, \TiFe~=~0.04~dex and \NiFe~=~0.05~dex.
SPADES then reanalyzed the spectrum another 50 times, but with
starting parameters randomly chosen following uniform distributions
centered on the true parameters and widths of $\pm$~500~K (\Teff),
$\pm$~0.5 (\logg) and $\pm$~0.4~dex (\FeH). The volume in the
parameters space considered for the initialization is larger,
i.e. conservative, with respect to the estimated performances of
SPADES-TE (see Sect.~\ref{TGMET}). Table~\ref{convergenceResiduals} 
presents the means and residuals of the parameters determined by SPADES
when initialized randomly minus those determined by SPADES when
initialized with the true parameters. Means and dispersions
are all low, showing that at S/N=50, SPADES is very weakly sensitive
to the initialization conditions.

\begin{table}[h!]
\caption{Means and dispersions of the residuals (SPADES initialized
randomly minus SPADES initialized with the true parameters) on the estimation
of the atmospheric parameters as well as the titanium and nickel over iron ratios.}             
\label{convergenceResiduals}
\centering
\begin{tabular}{l c c c c c}
         & \Teff & \logg   & \FeH     & \TiFe & \NiFe    \\
         & (K)   &         & (dex)    & (dex) & (dex)    \\ \hline
mean     &  1    & $-$0.01 & $-$0.007 & 0.002 & $-$0.007 \\
$\sigma$ &  3    &    0.02 &    0.007 & 0.004 &    0.007 \\
\end{tabular}
\end{table}


\section{Summary}
The automated stellar parameters determination software SPADES
was presented. It relies on a line by line comparison between
the studied spectrum and synthetic spectra. SPADES derives the
radial velocity, the effective temperature, the surface gravity,
and the iron and individual abundances.\\

SPADES internal systematic and random errors were assessed by
Monte Carlo method. For example, simulating Giraffe HR13 and HR14B setups from synthetic spectra
 for a star with $\Teff=5800$~K,
$\logg=4.5$, $\FeH=0.0$~dex and for a S/N
of 100, the stellar parameters were
recovered with no significant bias and with 1-$\sigma$ precisions
of 8 K for the temperature, 0.05 for the \logg, 0.009 for \FeH,
0.003 for \TiFe, and 0.01 for \NiFe. The external systematic errors
were evaluated with ground-based observed spectra and the convergence properties
were assessed by a Monte Carlo method.\\

Several evolutions of SPADES are planned. On the one hand, we aim to adapt
and test SPADES for a greater variety of spectral types and
luminosity classes. This includes addingare the micro-turbulence
to the parameters determined by SPADES. On the other hand,
we plant to implement additional diagnostics to determine the atmospheric
parameters, e.g. use the wings of the strong lines to derive
the surface gravity (for example Mg~Ib triplet in dwarfs or
infra-red Ca~II triplet in giants).


\begin{acknowledgements}
This work is financed by the CNRS (Centre National de la Recherche Scientifique) and the CNES (Centre National d'Etudes Spatiales).
We would like to thank R.~Kurucz for making publicly available the ATLAS9 and SYNTHE programs and F.~Castelli for providing the new
ODF tables. We are grateful to the referee for his/her very fruitful comments and suggestions.  
\end{acknowledgements}


\bibliography{posbic}
\bibliographystyle{aa}

\end{document}